\documentstyle [12pt]{article}
\newcommand{\agt}{{\lower 2pt\hbox{$>$} \atop \raise 2pt\hbox{$\sim$}}}
\newcommand{\alt}{{\lower 2pt\hbox{$<$} \atop \raise 2pt\hbox{$\sim$}}}

\begin{document}
\begin{titlepage}
\rightline{CMU-HEP94-17}
\rightline{DOE-ER/40682-71}
\leftline{hep-ph/9406306}
\leftline{May, 1994}
\vspace{3cm}
\begin{center}
{\LARGE\bf A MODEL FOR THE ORIGIN AND MECHANISMS OF CP VIOLATION}
\end{center}
\bigskip
\begin{center}
{\large\bf YUE-LIANG WU} \\
 Department of Physics, \\ Carnegie-Mellon University, \\
Pittsburgh, Pennsylvania 15213, U.S.A.
\end{center}
\vspace{3cm}
\begin{center}
{\bf Invited Talk at the 5th Conference on the Intersections of \\
Particle and Nuclear Physics \\ May 31-June 6, 1994, at St. Petersburg,
Florida \\ To appear in the  Proceedings, ed. S.J. Seestrom, (AIP, 1994)}
\end{center}

\end{titlepage}
\rightline{CMU-HEP94-17}
\rightline{DOE-ER/40682-71}
\leftline{hep-ph/9406306}
\leftline{May, 1994}
\bigskip
\begin{center}
{\Large\bf A Model for the \\ Origin and Mechanisms of CP Violation}
\end{center}
\bigskip
\begin{center}
{\bf Yue-Liang Wu \\
\small Department of Physics, Carnegie-Mellon University, \\
\small Pittsburgh, Pennsylvania 15213, U.S.A.}
\end{center}
\bigskip
\date{May, 1994}
\bigskip

\begin{abstract}
In this talk I will show that the two-Higgs doublet model  with vacuum
CP violation and approximate global $U(1)$ family symmetries
may provide one of the simplest and most attractive models for understanding
the origin and mechanisms of CP violation. It is shown that the mechanism of
spontaneous symmetry breaking provides not only a mechanism for generating
masses of the bosons and fermions, but also a mechanism for creating CP-phases
of the bosons and fermions, so that CP violation occurs, after spontaneous
symmetry breaking,  in all possible ways from a single CP phase of the vacuum
and is generally classified into four types of CP-violating mechanism.
A new type of CP-violating mechanism in the charged Higgs boson interactions
of the fermions is emphasized and can provide a consistent description for both
established
and reported CP-, P- and T-violating phenomena. Of particular importance is the
new source of CP violation for charged Higgs boson interactions that lead to
the value of $\epsilon'/\epsilon$ as large as $10^{-3}$ independend of
the CKM phase.
\end{abstract}

\newpage

\section{BASIC ASSUMPTIONS}

  The origin and mechanisms of CP violation have been investigated for
thirty years since the discovery of CP violation. In this talk I will present
the most recent work \cite{YLWU,WW} about new sources of CP violation.
I will show how the origin and mechanisms may be understood in the framework
of $SU(2)\times U(1)$ gauge theory with two-Higgs doublets. The model is
built based on the following two assumptions:

 (1) \   CP violation originates solely from a single CP-violating
phase of the vaccum.

  (2) \   The model exhibits Approximate Global U(1)
Family Symmetries (AGUFS) which act only on the fermions.

 The first assumption is introduced to understand the origin of CP violation.
 The second assumption is motivated from the smallness of the
Cabibbo-Kobayash-Maskawa quark mixing angles and the observed suppression of
the flavor-changing neutral currents.

\section{VACUUM CP VIOLATION}

In order to prevent the so-called domain-wall problem from arising explicitly
due to spontaneous CP violation, we observe the following fact that

   {\it In the gauge theories of spontaneous symmetry breaking,
CP violation can be originated solely  from the vacuum after
spontaneous symmetry breaking, even if CP symmetry is not good prior to the
symmetry breaking.} The {\it prerequisite condition} for such a statement is:
{\it CP nonconservation occurs only at one term in the
Higgs potential}.  This requirement actually results that
the vacuum must violate CP symmetry. In particular, this condition may be
simply realized by imposing an {\it universal rule} on the whole lagrangian.
That is, in a
renormalizable lagrangian all the  interactions with dimension-four  conserve
CP and only  interactions with dimension-two exhibit CP nonconservation.
It may also be naturally implemented through imposing some symmetries.
For convenience, we refer such a CP violation  as a
Vacuum CP Violation (VCPV).

Let me illustrate the simplest case with imposing the universal rule stated
above, the Higgs potential for the two Higgs doublets can be simply written
in the following general form
\begin{eqnarray}
V(\phi) & = & \lambda_{1}(\phi_{1}^{\dagger}\phi_{1}- \frac{1}{2} v_{1}^{2})^2
+ \lambda_{2}(\phi_{2}^{\dagger}\phi_{2} - \frac{1}{2} v_{1}^{2})^2 \nonumber
\\
& & + \lambda_{3}(\phi_{1}^{\dagger}\phi_{1} - \frac{1}{2} v_{1}^{2})
(\phi_{2}^{\dagger}\phi_{2} - \frac{1}{2} v_{2}^{2})
+ \lambda_{4}[(\phi_{1}^{\dagger}\phi_{1})(\phi_{2}^{\dagger}\phi_{2})
-(\phi_{1}^{\dagger}\phi_{2})(\phi_{2}^{\dagger}\phi_{1})]  \nonumber  \\
& &  + \frac{1}{2}\lambda_{5}(\phi_{1}^{\dagger}\phi_{2} +
\phi_{2}^{\dagger}\phi_{1} - v_{1}v_2 \cos\delta )^2
+ \lambda_{6}(\phi_{1}^{\dagger}\phi_{2}- \phi_{2}^{\dagger}\phi_{1} -
v_{1}v_{2} \sin\delta )^{2}  \\
& & + [\lambda_{7}(\phi_{1}^{\dagger}\phi_{1} - \frac{1}{2} v_{1}^{2})
+ \lambda_{8} (\phi_{2}^{\dagger}\phi_{2} - \frac{1}{2} v_{2}^{2})]
[\phi_{1}^{\dagger}\phi_{2} + \phi_{2}^{\dagger}\phi_{1} -
v_{1}v_2\cos\delta ]  \nonumber
\end{eqnarray}
where the $\lambda_i$ ($i=1, \cdots, 8$) are all real parameters.
If all the $\lambda_i$ are non-negative,  the minimum
of the potential then occurs at

\begin{equation}
<\phi_{1}^{0} > = \frac{v_1}{\sqrt{2}} e^{i\delta}\equiv \frac{v}{\sqrt{2}}
\cos\beta  e^{i\delta} \ ,  \qquad <\phi_{2}^{0} > = \frac{v_2}{\sqrt{2}}
\equiv \frac{v}{\sqrt{2}}\sin\beta
\end{equation}
It is clear that in the above potential
CP nonconservation can only occur through the vacuum, namely $\delta \neq 0$.
Obviously, such a CP violation appears to be explicit in the potential
when $\lambda_{6} \neq 0$, so that the domain-wall problem does not explicitly
arise. Note that if $\lambda_{6} = 0$, CP can still be violated
spontaneously\cite{TDL}. In general, one can also require one of other terms,
such as the term $\lambda_{5}$ or $\lambda_{7}$ or $\lambda_{8}$,
to be complex in the most general potential.

\section{APPROXIMATE GLOBAL U(1) FAMILY SYMMETRY}

To see why assuming approximate global U(1) family symmetries (AGUFS) rather
than approximate flavor symmetries, we start with a general Yukawa interaction

\begin{equation}
L_{Y} = \bar{q}_{L}\Gamma^{a}_{D} D_{R}\phi_{a} + \bar{q}_{L}\Gamma^{a}_{U}
U_{R}\bar{\phi}_{a} + \bar{l}_{L}\Gamma^{a}_{E} E_{R}\phi_{a} + H.C.
\end{equation}
where $q^{i}_{L}$, $l^{i}_{L}$ and $\phi_{a}$ are $SU(2)_{L}$ doublet quarks,
leptons and Higgs bosons, while $U^{i}_{R}$, $D^{i}_{R}$ and $E^{i}_{R}$
are $SU(2)_{L}$ singlets. $i = 1,\cdots , n_{F}$ is a family label and
$a = 1, \cdots , n_{H}$ is a Higgs doublet label. $\Gamma^{a}_{F}$
($F= U, D, E$) are the arbitrary real Yukawa coupling matrices and
parameterized
in the following general form

\begin{equation}
\Gamma^{a}_{F}  =  O_{L}^{F} \sum_{i,j=1}^{n_{F}}\{ \omega_{i} (g_{a}^{F_{i}}
\delta_{ij}  + \zeta_{F} \sqrt{g^{F_{i}}} S_{a}^{F}\sqrt{g^{F_{j}}} )
\omega_{j} \} (O_{R}^{F})^{T}
\end{equation}
where the first term in the brackets is a diagonal matrix and the second term
is an off-diagonal matrix. To facilitate a comparison between the elements of
the two parts,  $g^{F_{i}}$ is taken to be $g^{F_{i}} = | \sum_{a}
g^{F_{i}}_{a} \hat{v}_{a} | / (\sum_{a} |\hat{v}_{a}|^{2})^{\frac{1}{2}}$ .
$\{\omega_{i}, i=1, \cdots,n_{F}\}$ the set of diagonalized projection matrices
$(\omega_{i})_{jj'} = \delta_{ji}\delta_{j'i}$. $\hat{v}_{a}\equiv
<\phi^{0}_{a}>$ ($a=1, \cdots , n_{H}$) are Vacuum Expectation Values (VEV's)
which will develop from the Higgs bosons after spontaneous symmetry breaking.
$g^{F_{i}}_{a}$ are the arbitrary real Yukawa coupling constants.
$S_{a}^{F}$ ($a\neq n_{H}$) are the arbitrary off-diagonal real matrices
(i.e. $(S_{a}^{F})_{ii} \equiv 0$). By convention, we choose
$S_{a}^{F}=0$ for $a=n_{H}$ to eliminate the dependent parameters.
$\zeta_{F}$ is a conventional parameter introduced to scale the
off-diagonal matrix elements so that $(S_{1}^{F})_{12}\equiv 1$.
$(S_{a}^{F})_{ij}$ are expected to be of order unity (some elements of
$S_{a}^{F}$ may be off by a factor of 2 or more).  $O_{L,R}^{F}$ are the
arbitrary orthogonal matrices. Note that the above parameterization is general
but very useful and powerful for our purposes in analysing various
interesting physical phenomena.   In general, one can always choose,
by a redifinition of the  fermions, a basis so that $O_{L}^{F}=O_{R}^{F}\equiv
O^{F}$ (and $O^{U} = 1$ or $O^{D} = 1$ if it needs) as well as $O^{E}=1$
since the neutrinos are considered to be massless in this model

   In the gauge interactions of the fermions, there exist large global
(chiral) {\it flavor} symmetries $U(3)_{L}^{q}\times U(3)_{R}^{U}\times
U(3)_{R}^{D}\times U(3)_{L}^{l}\times U(3)_{R}^{E}$. These symmetries
have to be broken down to the global (vector-like) {\it family} symmetries as
long as the scalar interactions of the fermions are introduced to the model.
This is because the scalar interactions of the fermions change the
chirality of the fermions. The resulted global family symmetries depend on
the physical considerations. To see this, let us consider several special
cases: Suppose
that the masses of all the up-type quarks, and also the down-type quarks as
well as the leptons, are equal, and there are no quark mixings and also
no flavor-changing neutral scalar interactions, i.e., $g_{a}^{F_{i}} =
g_{a}^{F_{j}} $, $O^{U}= O^{D}=1$ and $\zeta_{F} = 0$, then the above large
global (chiral) {\it flavor} symmetries will be broken down to the large global
(vector-like) {\it family} symmetries $U(3)^{q}\times U(3)^{l}$. This symmetry
violation must be large as the top quark is heavy and its Yukawa coupling is of
order unity.   It is known that all the fermions have
different masses, suppose that the quark mixings and the flavor-changing
neutral scalar interactions are still zero, thus the large global family
symmetries are further broken down to the global U(1) family symmetries
$U(1)_{1}^{q}\times U(1)_{2}^{q}\times U(1)_{3}^{q}\times U(1)_{1}^{l}\times
U(1)_{2}^{l}\times U(1)_{3}^{l}$, one for each family.
Due to the large mass splitting among the fermions, for instance,
$m_{t}/m_{u} \sim 3\times 10^{4}$, the symmetry violation
should also be large. In the realistic case, the quarks do mix and there
exist flavor-changing neutral currents, therefore the global U(1) family
symmetries must be completely broken down.  Nevertheless, the CKM matrix
is known to deviate only slightly from unity and the observed flavor-changing
neutral currents are strongly suppressed, so that
at the electroweak scale any successful model will exhibit approximate
global $U(1)$ family symmetries (AGUFS). In other words, the AGUFS are
sufficient  for a natural suppression of family-changing currents
(for both charged and neutral currents). Explicitly, AGUFS indicate that

\begin{equation}
(O^{F})_{ij}^{2} \ll 1 \  , \qquad i\neq j \ ; \qquad  \zeta_{F}^{2}  \ll 1 \
{}.
\end{equation}
where $O^{F}$ describe the AGUFS in the charged currents and $\zeta_{F}$
mainly characterizes the AGUFS in the neutral currents.

It should be noted that our considerations are unlike the Hall and
Weinberg's \cite{HW} approximate global U(1) {\it flavor} symmetries, i.e.,
one for each fermion type. This is because we do not
specify the violation  of global U(1) symmetries to the mass parameters
in our approximate global $U(1)$ {\it family} symmetries (AGUFS) ( one for
each family).

\section{CP-VIOLATING INTERACTIONS}

 With the above general analyses and assumptions,
 let me now present a detailed description for the model with vacuum CP
violation (VCPV) and approximate global U(1) family symmetries
(AGUFS).  The physical interactions  are usually given in the mass basis of
the particles. For the simplest two-Higgs doublet model, the
physical basis  after spontaneous symmetry breaking  is defined
through $f_L = (O_{L}^{F}V_{L}^{f})^{\dagger}F_L$ and $ f_R = (O_{R}^{F}P^f
V_{R}^{f})^{\dagger}F_R $ with $V_{L,R}^{f}$ being unitary matrices and
introduced to diagonalize the mass matrices
\begin{equation}
(V_{L}^{f})^{\dagger}(\sum_{i}m_{f_{i}}^{o}\omega_{i} + \zeta_{F} c_{\beta}
\sum_{i,j} \sqrt{m_{f_{i}}^{o}} \omega_{i} S_{1}^{F} \omega_{j}
\sqrt{m_{f_{j}}^{o}} e^{i\sigma_{f}(\delta - \delta_{f_{j}})}) V_{R}^{f} =
\sum_{i} m_{f_{i}}\omega_{i}
\end{equation}
with $m_{f_{i}}$ the masses of the physical states $f_{i}= u_i, d_i, e_i$.
Where $m_{f_{i}}^{o}$ and $\delta_{f_{i}}$ are defined via
\begin{equation}
(c_{\beta} g_{1}^{F_{i}}e^{i\sigma_{f}\delta} +
s_{\beta} g_{2}^{F_{i}})v \equiv \sqrt{2}m_{f_{i}}^{o} e^{i\sigma_{f}
\delta_{f_{i}}}
\end{equation}
 with $v^2 = v_{1}^{2} + v_{2}^{2}= (\sqrt{2}G_{F})^{-1}$,
$ c_{\beta}\equiv \cos\beta = v_1/v$ and $s_{\beta}\equiv \sin\beta
= v_2/v$. $P^{f}_{ij} = e^{i\sigma_{f} \delta_{f_{i}}}\delta_{ij}$,
with $\sigma_{f} =+$, for $f= d, e$, and  $\sigma_{f} = - $, for $f = u$.

  For convenience, we fix the phase convention by writing

\[ V_{L,R}^{f} \equiv  1 + \zeta_{F}c_{\beta} T_{L,R}^{f} \]

 In a good approximation, to the first order in $\zeta_{F}$ and the lowest
order in $m_{f_{i}}/m_{f_{j}}$ with $i < j$, we find
that $m_{f_{i}}^{2} \simeq (m_{f_{i}}^{o})^{2} + O(\zeta^{2}_{F})$ and
for $i < j$
\begin{eqnarray}
(T_{L}^{f})_{ij} & \simeq & - (T_{L}^{f})_{ji}^{\ast} \simeq
\sqrt{\frac{m_{f_{i}}}{m_{f_{j}}}} (S_{1}^{F})_{ij} e^{-i\sigma_{f} (\delta
- \delta_{f_{j}})} + O((\frac{m_{f_{i}}}{m_{f_{j}}})^{3/2}, \zeta_{F}) \\
(T_{R}^{f})_{ij} & \simeq & - (T_{R}^{f})_{ji}^{\ast} \simeq
\sqrt{\frac{m_{f_{i}}}{m_{f_{j}}}} (S_{1}^{F})_{ji} e^{-i\sigma_{f} (\delta
- \delta_{f_{j}})} + O((\frac{m_{f_{i}}}{m_{f_{j}}})^{3/2}, \zeta_{F})
\end{eqnarray}

 With this phase convention, the CKM matrix $V$ has the following form
\begin{equation}
V   = (V_{L}^{U})^{\dagger} (O_{L}^{U})^{T} O_{L}^{D} V_{L}^{D} \equiv
V^{o} + V'
\end{equation}
where $V^{o} \equiv  (O_{L}^{U})^{T} O_{L}^{D}$ is a real matrix and
$V' \simeq \zeta_{D}c_{\beta} [V^{o} T_{L}^{d}] + \zeta_{U} c_{\beta}
[V^{o T}T_{L}^{u}]^{\dagger}$ is a complex matrix. It is obvious that
when $\zeta_{F}\rightarrow 0$, then $V' \rightarrow 0$, the CKM matrix
is almost described by the real matrix $V^{o}$.

{\it  The scalar interactions of the fermions} read in the physical basis
\begin{eqnarray}
L_{Y}^{I} & = & (2\sqrt{2}G_{F})^{1/2}\sum_{i,j,j'}^{3}H^{+} \{
\bar{u}_{L}^{i} V_{ij'} (m_{d_{j'}}\xi_{d_{j'}}\delta_{j'j} +
\frac{\zeta_{D}}{s_{\beta}}
\sqrt{m_{d_{j'}}m_{d_{j}}} S_{j'j}^{d}) d^{j}_{R}
- H^{-} \bar{d}_{L}^{i} V_{ij'}^{\dagger}
\nonumber \\
& & \cdot (m_{u_{j'}}\xi_{u_{j'}}\delta_{j'j} + \frac{\zeta_{U}}{s_{\beta}}
\sqrt{m_{u_{j'}}m_{u_{j}}} S_{j'j}^{u}) u^{j}_{R}
+ H^{+} \bar{\nu}_{L}^{i} (m_{e_{i}}\xi_{e_{i}}\delta_{ij} +
\frac{\zeta_{E}}{s_{\beta}}
\sqrt{m_{e_{i}}m_{e_{j}}} S_{ij}^{e}) e^{j}_{R}\} \nonumber \\
& & + (\sqrt{2}G_{F})^{1/2}\sum_{i,j}^{3} \sum_{k}^{3} H^{0}_{k}
\{ \bar{u}^{i}_{L}
(m_{u_{i}} \eta_{u_{i}}^{(k)} \delta_{ij} + \frac{\zeta_{U}}{s_{\beta}}
\sqrt{m_{u_{i}}m_{u_{j}}}S_{k,ij}^{u}) u^{j}_{R} + \bar{d}^{i}_{L}
(m_{d_{i}} \eta_{d_{i}}^{(k)} \delta_{ij}  \nonumber \\
 & & + \frac{\zeta_{D}}{s_{\beta}} \sqrt{m_{d_{i}}m_{d_{j}}}
S_{k,ij}^{d}) d^{j}_{R}
+ \bar{e}^{i}_{L}(m_{e_{i}} \eta_{e_{i}}^{(k)} \delta_{ij} +
\frac{\zeta_{E}}{s_{\beta}}\sqrt{m_{e_{i}}m_{e_{j}}} S_{k,ij}^{e}) e^{j}_{R}\}
 + H.C.
\end{eqnarray}
with (in the above good approximations)
\begin{eqnarray}
& & \xi_{f_{i}} \simeq \frac{\sin\delta_{f_{i}}}
{s_{\beta}c_{\beta}\sin\delta}e^{i \sigma_{f}(\delta - \delta_{f_{i}})}
 - \cot\beta \  ,  \\
& & S^{f}_{ij} \simeq (e^{i\sigma_{f}(\delta - \delta_{f_{j}})}
- \frac{\sin\delta_{f_{j}}}{\sin\delta} ) (S_{1}^{F})_{ij} \  , \qquad
S^{f}_{ji} \simeq   (e^{i\sigma_{f}(\delta - \delta_{f_{i}})}
- \frac{\sin\delta_{f_{j}}}{\sin\delta} ) (S_{1}^{F})_{ji} \  . \\
& & \eta_{f_{i}}^{(k)}  = O_{2k}^{H}  +  (O_{1k}^{H} + i \sigma_{f} O_{3k}^{H})
\xi_{f_{i}} \ ; \qquad S_{k,ij}^{f} = (O_{1k}^{H} + i \sigma_{f}
O_{3k}^{H} )S^{f}_{ij}\ .
\end{eqnarray}
where $O_{ij}^{H}$ is the orthogonal matrix introduced to redefine the
three neutral scalars $ \hat{H}_{k}^{0} \equiv (R, \hat{H}^{0}, I)$
into their mass eigenstates $H_{k}^{0} \equiv (h, H, A)$, i.e. $\hat{H}_{k}^{0}
 =  O^{H}_{kl} H_{l}^{0}$ with $(R+iI)/\sqrt{2} =
s_{\beta}\phi^{0}_{1}e^{-i\delta} - c_{\beta} \phi^{0}_{2}$ and
$(v+\hat{H}^{0} +iG^{0})/\sqrt{2} = c_{\beta}\phi^{0}_{1}
e^{-i\delta} + s_{\beta} \phi^{0}_{2} $.
Here $H_{2}^{0}\equiv H^{0}$ plays the role of the Higgs boson in the
standard model. $ H^{\pm}$ are the charged scalar pair with $ H^{\pm} =
s_{\beta}\phi^{\pm}_{1}e^{-i\delta} - c_{\beta} \phi^{\pm}_{2}$.

\section{ORIGIN AND MECHANISMS OF CP VIOLATION}

  By explicitly giving the new CP-violating interactions, we see that
if the relative phase between the two VEV's is
nonzero,  each fermion $f_{i}$ and neutral scalar $H_{k}^{0}$  are then
characterized not only by their physical masses $m_{f_{i}}$ and
$m_{H_{k}^{0}}$   but also by their physical phases

\begin{equation}
\delta_{f_{i}} \equiv arg ((c_{\beta} g_{1}^{F_{i}}e^{i\sigma_{f}\delta} +
s_{\beta} g_{2}^{F_{i}})v) \sigma_{f}  \qquad
 \delta_{H_{k}^{0}} \equiv arg(O_{1k}^{H} + i \sigma_{f} O_{3k}^{H})
\end{equation}
This implies that the Higgs mechanism
provides not only a mechanism for generating the masses of the bosons and
fermions, but also a mechanism for creating CP-phases of the bosons and
fermions.

  With the above sources of CP violation, CP violation occurs in all
possible ways.  All these vacuum-induced CP
violations can be classified  into four
types of mechanism according to their  interactions:

 {\bf Type-I.} \  The new type of CP-violating mechanism
which arises from the induced complex diagonal Yukawa couplings $\xi_{f_{i}}$.
Such a CP violation can occur through both charged- and neutral-scalar
exchanges.

{\bf Type-II.}\  Flavor-Changing Superweak -type  mechanism due to the
flavor-changing scalar interactions.
This type of mechanism also occurs through both charged- and neutral-scalar
exchanges and is described by the complex coupling matrices $S_{ij}^{f}$
in this model.

 {\bf Type-III.}\  The induced KM-type CP-violating mechanism  which
is characterized in this model by the complex parameters $\zeta_{F} T_{L}^{f}$
and occurs in the both charged gauge boson and  charged scalar interactions
of the quarks.

 {\bf Type-IV.}\  The Scalar-Pseudoscalar Mixing  mechanism
which is described by the mixing matrix $O_{kl}^{H}$ and the phases
$arg(O_{1k}^{H}+i\sigma_{f}O_{3k}^{H})$. This type of CP
violation appears in the purely bosonic interactions and also
in the neutral scalar interactions of the fermions in this
model. In general, such kind of scalar mixing mechanism can also occur in
the charged scalar interactions of the fermions when there exist more than
two charged scalars, for example, the Weinberg three-Higgs doublet model.

\section{PHENOMENOLOGY OF CP VIOLATION}

   After understanding the origin and mechanisms of CP violation, let me
now briefly discuss their main features and summarize the most interesting
physical phenomena (for systematic analyses and detailed calculations see
\cite{YLWU}).

 1) \  Without making any  additional assumptions, $m_{f_{i}}$, $V_{ij}$,
$\delta_{f_{i}}$ (or $\xi_{f_{i}}$), $\delta$, $\tan\beta$, $\zeta_{F}$,
$(S_{1}^{F})_{ij}$ (or $S_{ij}^{f}$), $m_{H_{k}^{0}}$, $m_{H^{+}}$ and
$O_{kl}^{H}$ are in principle all the free parameters and will be determined
only by the experiments. Nevertheless, from the AGUFS,  we can  draw
the general features that $V_{ij}^{2} \ll 1$ for $i \neq j$ and
$\zeta_{F}^{2} \ll 1$. The $m_{f_{i}}$ and $V_{ij}$ already appear in the
standard model and have been extensively investigated. For the other
parameters, it is expected that  $(S_{1}^{F})_{ij}$ are of order unity.
Moreover, in order to have the flavor-changing neutral scalar interactions
(FCNSI) be suppressed manifestly, it is in favor
of having  $\tan\beta > 1$ and  $|\sin\delta_{f_{j}}/\sin\delta | \leq 1 $.
The diagonal Yukawa couplings $\eta_{f_{i}}^{(k)}$ or $\xi_{f_{i}}$ can be,
for the light fermions,  much larger than one (the corresponding couplings
are equal to one in the standard model)  and may, of course,
also be smaller than one (this case appears to
occur for heavy top quark). Nevertheless, the former case is more
attractive since it will result in significant interesting phenomenological
effects.  The most interesting choice for large $\xi_{f_{i}}$ is
$\tan\beta \gg 1$ since it favors the suppression of the flavor-changing
neutral scale interactions (FCNSI).

  We would like to point out that the phase convention of the CKM matrix is
nontrivial in this model due to the existence of the flavor-changing
neutral scalar interactions (FCNSI).  Therefore, before discussing
contributions to CP violating observables from various mechanisms, one should
first specify the phase convention.

2)\ From the established $K^{0}-\bar{K}^{0}$ and $B^{0}-\bar{B}^{0}$ mixings,
we obtain that $\zeta_{D}/s_{\beta} < 10^{-3} m_{H_{k}^{0}}/GeV$.
{}From the current experimental bound of the $D^{0}-\bar{D}^{0}$ mixing,
i.e. $\Delta m_{D} < 1.3 \times 10^{-4}$ eV, we have
$\zeta_{U}/s_{\beta} < 3\times 10^{-3} m_{H_{k}^{0}}/GeV$. Note that
this can only be regarded as an order-of-magnitude estimation since in
obtaining
these values we have used the vacuum insertion approximation for the
evaluation of the hadronic matrix elements. From the established CP-violating
parameter $\epsilon$, it requires either to fine-tune the parameters
$\delta_{d}$ , $\delta_{s}$, $arg(O_{1k}^{H} + i O_{3k}^{H})$,
$(S_{1}^{D})_{12}$ {\it etc.}, so that the effective CP-phase is of order
$10^{-2}$ or to choose $\zeta_{D}/s_{\beta} < 10^{-4} m_{H_{k}^{0}}/GeV$.
For the latter case, the effective CP-violating phases are indeed generically
of order unity.

 3) \  To see how the various mechanisms play the role on CP violation and
provide interesting physical phenomena, let us consider the
following three cases:

(i) when $\zeta_{F}/s_{\beta}\ll 1$,  $(S_{1}^{F})_{ij} = O(1)$,
 $m_{H^{+}} \leq v =246$ GeV and  $|\xi_{f_{i}}| \gg 1$ ($f_{i}\neq t$),
it is obvious that only the new type of CP-violating mechanism (type-I) plays
an important role.  The effects from the flavor-changing superweak -type
and KM-type mechanisms are negligible.

 In this case, the CP-violating parameter $\epsilon$ can be
fitted from the contributions of the short-distance box graph with charged
scalar exchanges and the long-distance dispersive
through the $\pi$, $\eta$ and $\eta'$. This is easily
achieved in our model through choosing appropriate parameters
$Im(\xi_{s}\xi_{c})^{2}$ and $Im(\xi_{s}\xi_{c})$  for a
given mass $m_{H^{+}}$.  The ratio $\epsilon'/\epsilon$ could be of
order $10^{-3}$ from both the short-distance contribution at tree level through
the charged scalar exchange and the long-distance contribution induced by
the charged scalar penguin diagram. The short-distance contribution to
$\epsilon'/\epsilon$ depends on the parameters $Im(\xi_{s}\xi_{d}^{\ast})$
and $Im(\xi_{s}\xi_{u})$.
The neutron EDM $d_{n}$ can be consistently accommodated by choosing other
independent parameters, such as $Im(\xi_{d}\xi_{c})$ and $Im (\eta_{d}^{(k)}-
\eta_{u}^{(k)})^{2}$ for the one-loop contribution with charged scalar and
neutral scalar exchanges respectively, and $Im(\xi_{t}\xi_{b})$ and
$Im(\eta_{t}^{(k)})^{2}$ for the Weinberg gluonic operator contribution
with charged scalar and neutral scalar exchanges respectively, as well as
$Im(\xi_{t}\xi_{q})$ ($q=d,s,u$) for the quark gluonic chromo-EDM.
Its value is not far below to the present experimental bound and
could be of order $10^{-26}$ e cm.
The electon EDM $d_{e}$ from Barr-Zee two-loop mechanism is expected
to be in the present experimental sensitivity for appropriate values of
$Im(\eta_{t}^{(k)}\eta_{e}^{(k)})$ and $Im(\eta_{t}^{(k)}\eta_{e}^{(k) \ast})$
as well as $O_{2k}^{H} Im\eta_{e}^{(k)}$.  CP violation in the hyperon
decay can also be significant in this case. For example, the CP asymmetry
observable $A(\Xi^{-} \rightarrow \Lambda \pi^{-}) \sim 10^{-4}$ which is
comparable to or even larger than in the standard model.
Direct CP violation in B-meson decay is, however, small in this limit case
since the phases of the mixing mass matrix and the amplitudes are suppressed
either by the mass ratio $m_{b}m_{d}/m_{t}^{2}$ or by $\zeta_{F}$.
Nevertheless, T-odd and CP-odd triple-product correlations could be
relatively large in the inclusive
and exclusive semileptonic decays of B-meson into the $\tau$ leptons.

 In addition, by including the new contributions to the neutral meson mixings
from the box diagrams with charged-scalar exchange,
the mass difference $\Delta m_{K}$ could be reproduced by a purely
short-distance analysis when $|\xi_{c}| \gg 1$.
For example, for $B_{K} = 0.7$, and $m_{H^{+}} = 100$ GeV, it needs
$|\xi_{c}|\simeq 13$. When $|\xi_{t}|\sim 1$, $B^{0}-\bar{B}^{0}$ and
$B_{s}^{0}-\bar{B}_{s}^{0}$ mixings could also receive a contribution as
large as in the standard model.

  For more explicit, let me present a simple numerical example with the
following scenario:

\[ \zeta_{F} < 10^{-3}, \qquad  \tan\beta = \frac{v_{2}}{v_{1}} \gg 1,
\qquad |\frac{\sin\delta_{f_{i}}}{\sin\delta}| = O(1) \]
with $f_{i} = u,d,s,c$, thus

\[ \xi_{f_{i}} \rightarrow \frac{\sin\delta_{f_{i}}}{\sin\delta}
e^{i\sigma_{f}(\delta - \delta_{f_{i}})} \tan \beta,
\qquad |\xi_{f_{i}}| \sim \tan\beta \]
This can be easily achieved by taking

\[ \delta \sim \frac{\pi}{3}\ , \qquad
\delta_{s} \sim \delta_{d} \sim \frac{\pi}{4} \qquad
\delta_{u}\sim \frac{\pi}{2} + \frac{\pi}{4}, \qquad
\delta_{c}\sim \frac{\pi}{2} + \frac{\pi}{4} - \frac{\pi}{50} \]
which lead to

\[ Im (\xi_{c}\xi_{s}) \sim Im (\xi_{u}\xi_{s}) \sim \tan^{2}\beta \ ,
 \qquad Im (\xi_{c}\xi_{s})^{2} \sim \frac{1}{8}\tan^{4}\beta,
\qquad Im (\xi_{d}^{\ast}\xi_{s}) \sim 0 \]
as long as

\[ \tan\beta \sim 10\  (m_{H^{+}}/200GeV) \]
we have
\[ |\epsilon| = 2.27\times 10^{-3} \qquad
\frac{\epsilon'}{\epsilon} \sim (1-3) \times 10^{-3} \]
\[ \phi (\psi K_{S}) \sim 10^{-2}, \qquad A(\Xi^{-} \rightarrow \Lambda
\pi^{-}) \sim 10^{-4}, \qquad  d_{n} \sim    10^{-26} \mbox{e cm} \]
 This simple numerical example shows that the new type of CP-violating
mechanism
could be indeed significant.

(ii) when $10^{-4}m_{H_{k}^{0}}/GeV < \zeta_{D}/s_{\beta} < 0.1$,
$\zeta_{U}/s_{\beta} < 0.3$ for $m_{t} = 175$ GeV,
$(S_{1}^{F})_{ij} = O(1)$ and $|\xi_{f_{i}}| < 1$, both the new type of
CP-violating mechanism and the induced KM-type mechanism become less important
and the parameter $\epsilon$ is then accounted for by the flavor-changing
superweak-type mechanism
(type-II) together with the type-IV. If the CP-violating
phases are indeed generically of order unity, thus the ratio
$\epsilon'/\epsilon$ becomes unobservably small ($\sim 10^{-6}$).
In this case, its effects in the B-system are also small. Whereas if
$B^{0}-\bar{B}^{0}$ and $B_{s}^{0}-\bar{B}_{s}^{0}$ mixings receive large
contributions from the flavor-changing neutral scalar interactions, any
value of CP asymmetry in B-system can occur \cite{SW}.

(iii) when $\zeta_{D}/s_{\beta} \sim 0.2$ and $\zeta_{U}/s_{\beta} \sim 0.6$
for $m_{t}=175$ GeV, $c_{\beta}\sim s_{\beta}$, $|\xi_{i}| \sim 1$ and
$m_{H^{0}_{k}} \gg v =246$ GeV, i.e., neutral scalars are very heavy, then
the CP-violating mechanism is governed by the induced KM-type mechanism.
But it could be still different from the standard KM-model if the charged
scalar is not so heavy, this is because the new contributions from diagrams
with charged-scalar exchange can be significant. Therefore only when
the charged scalar also become very heavy, the induced KM-type mechanism
then coincides with  the standard KM-model which have been extensively
studied \cite{WLW}.

  In general, four types of CP-violating mechanism may simultaneously
play an important role for the CP-violating phenomena.

\section{CONCLUSIONS}

  In conclusion, it is seen that precisely measuring the direct CP violation
in kaon (and hyperon) decays and the direct CP violation in B-system as well as
the EDM's of the electron and the neutron are very important for clarifying
the origin and mechanisms of CP violation. For instance, if the direct CP
violation in kaon decay is large and confirmed to be of order $10^{-3}$, the
neutron and electron EDMs are also in the present observable level,
while direct CP violation in B-system is smaller than the standard model
prediction, it then clearly indicates that the new type of CP-violating
mechanism will be important.

  Based on the assumption of the approximate global U(1) family symmetries
(AGUFS), the mass of the
scalars could be less constrained from the indirect experimental data, such as
neutral meson mixings and inclusive bottom quark decays $b\rightarrow s
\gamma$. Searching for
these exotic scalars  is worthwhile at both $e^{+}e^{-}$ and hadron colliders.
It is believed that the mechanisms of CP violation discussed in this
model should also play an important role for understanding the baryogenesis
at the electroweak scale. In particular, its requirement for
relatively light Higgs bosons in order to avoid the wash out problem
is in favor of our model. All these
considerations and new features suggest that if one Higgs doublet is necessary
for the generation of
the masses of the bosons and fermions, then two Higgs doublets are needed
for the origin and phenomenology of CP violation and also for baryogenesis at
the electroweak scale.

This work was supported by DOE grant \# DE-FG02-91ER40682.

\end{document}